# Spin Control of Drifting Electrons using Local Nuclear Polarization in Ferromagnet/Semiconductor Heterostructures

M.E. Nowakowski[1], G.D. Fuchs[1], S. Mack[1], N. Samarth[2], D.D. Awschalom[1]*

[1.] *Center for Spintronics and Quantum Computation, University of California, Santa Barbara, CA 93106, USA*

[2.] *Department of Physics and Materials Research Institute, The Pennsylvania State University, University Park, PA 16802*

**Abstract**

We demonstrate methods to locally control the spin rotation of moving electrons in a GaAs channel. The Larmor frequency of optically-injected spins is modulated when the spins are dragged through a region of spin-polarized nuclei created at a MnAs/GaAs interface. The effective field created by the nuclei is controlled either optically or electrically using the ferromagnetic proximity polarization effect. Spin rotation is also tuned by controlling the carrier traverse time through the polarized region. We demonstrate coherent spin rotations exceeding $4\pi$ radians during transport.

The ability to manipulate the state of electron and nuclear spins is a critical tool in spin-based quantum science. Control of spins typically requires pulsed optical or microwave fields to manipulate stationary spin ensembles or single spin centers [1-6]. Using these techniques and judiciously chosen pulse sequences, it is possible to prepare confined electrons into any coherent superposition of spin-½ eigenstates [7]. In a transport geometry, applied voltages cause electrons to drift, and the ability to locally control the spin state of these moving charges [8] is another fundamental challenge with potential technological impact for quantum information processing. Implementing a quantum gate operation during transport requires the ability to initialize, coherently manipulate during drift, and read-out electron spins within the coherence time [9]. Conduction electron spin ensembles in n-type GaAs are promising candidates, as they can have coherence times above 100 ns at low temperatures [10], can be optically [11] or electrically [12] initialized, transported over distances exceeding 100 μm [11], and read-out both optically [11] and electrically [13]. Previous experiments in lateral geometries have demonstrated non-local electrical, magnetic, and spin-orbit field control of drifting spins in GaAs and InGaAs quantum wells [12-15] and magnetic control in Si [16] and graphene [17]. Additionally, local optical control of the spin polarization amplitude has been demonstrated in a GaAs ring interferometer structure [18] using dynamic nuclear polarization (DNP).

We performed a spatially-localized quantum gate operation on an ensemble of moving spins using a prepared region of polarized GaAs nuclear spins in a lateral transport structure. Optically-injected electron spin ensembles [11] are dragged by an applied voltage through the polarized nuclear region [19], emerging on the other side with a rotated spin state. The nuclear spins are polarized at the interface of a lithographically-patterned epitaxial MnAs ferromagnet via the ferromagnetic proximity polarization (FPP) effect [20, 21]. This effect can be activated either optically [20] or electrically [22, 23] in the channel and, in the presence of a 0.2 T magnetic field at low temperatures, has been shown to create an additional localized effective magnetic field under the MnAs up to 0.4 T [19]. In this region, the moving spins precess faster than they would in a bare channel, and as they emerge, optical measurements indicate they have acquired additional spin rotation about the z-axis defined by the magnetic field. We tune the angle of rotation by controlling the local nuclear field strength and interaction time between the polarized nuclei and moving electrons. In our experiment, the rotation angle of the moving spins is modified by more than 4π radians along the Bloch sphere in 7 ns over a distance of 30 microns.

Samples were fabricated from a layer structure of 25 nm type-A MnAs/1 μm Si-doped n-GaAs (n=$1\times10^{17}$cm$^{-3}$)/250 nm Al$_{0.45}$Ga$_{0.55}$As/semi-insulating (001) GaAs substrate grown by molecular beam epitaxy. Similar samples with lower Si doping ($3\times10^{16}$ and $7\times10^{16}$ cm$^{-3}$) demonstrated similar results but are not shown. The ferromagnetic MnAs layer exhibited square hysteresis with a coercivity of 2000 Oe at 10 K along the easy axis, in the <110> direction of the GaAs substrate. Using standard lithographic methods, we fabricated electrically-isolated 10 μm wide GaAs channels with 30 μm long MnAs islands above the center of the channel [Fig. 1(c)]. We made ohmic contact to the GaAs channel with a two-point resistance of 5 kΩ.

Electron spin dynamics were measured using time-resolved Kerr rotation (TRKR) in the Voigt geometry [24] with the sample growth axis parallel to the optical axis. A saturating external field, $B_{ext}$, of 0.2850 T was applied along the MnAs easy axis, parallel to the channel, as shown in Fig. 1(c). Optical excitation and read out were performed using a mode-locked titanium sapphire laser producing ~150 fs pulses at a repetition rate of 76 MHz. Its wavelength was tuned to 807 nm to excite carriers above the GaAs band gap. A circularly-polarized, 670 µW pump beam was used to inject spin-polarized electrons with an initial spin projection parallel to the optical axis. After a time delay, $\Delta t$, a linearly-polarized, 90 µW beam was used to probe the electron spin component along the optical axis via the magneto-optic Kerr effect. Both pump and probe beams were focused to a 5 µm spot size and were modulated at 50 kHz and 500 Hz, respectively, for lock-in detection, while $\Delta t$ was controlled by a mechanical delay line. Measurements were conducted at $T$ = 8 K.

The pump and probe beams were focused on the GaAs channel to the right and left of the MnAs island, respectively [Fig. 1(c)]. A voltage bias was applied to the channel causing the injected spins to drift from the pump spot to the probe spot. The 13.15 ns repetition time of the laser exceeds the 6 ns spin lifetime of our samples; therefore, the contrast in the probe signal is due solely to a single initialization pulse.

For nuclear spin polarization, a circularly-polarized, 5 mW imprinting beam, derived from the same laser, was focused onto the MnAs island with no bias applied as shown in Fig. 1(a). Excited electron spins spontaneously polarize along the ferromagnetic magnetization direction via the FPP effect [25]. The injected spins then transfer their angular momentum to the nuclear spins via the hyperfine interaction. This aligns the nuclear spins preferentially along the MnAs magnetization direction [20] adding a spatially-isolated [19] effective magnetic field, $B_{nuc}$, to the external field, $B_{ext}$, under the MnAs which we measured to be near 0.4 T. After imprinting for 20 minutes to saturate the nuclear spin polarization, the imprinting beam was blocked. We then turned on the channel voltage dragging the injected electron spins through the polarized nuclear region. Their spin projection was measured by Kerr rotation as they emerged on the other side.

We first characterize the drift and diffusion characteristics of a 10 μm wide channel without a MnAs island. We measured TRKR vs. delay and pump/probe separation ($\Delta x$) for channel currents ($i_{channel}$) between 0.5 and 2 mA. Figure 2(a) is a density plot of KR vs. delay and pump/probe separation for a channel current of 1.0 mA. The data are in good agreement with the one-dimensional spin drift-diffusion equation [12, 14]

$$s(x,t) = \frac{S_0}{\sqrt{4\pi D \Delta t}} \exp\left(\frac{-(x-v_d \Delta t)^2}{4D\Delta t}\right) \exp\left(\frac{-\Delta t}{T_2^*}\right) \cos(\omega_L \Delta t), \qquad (1)$$

where $S_0$ is the amplitude, $D$ is the spin diffusion constant, $v_d$ is the spin drift velocity, $x$ is the position of injection, $T_2^*$ is the spin lifetime, and $\omega_L = g\mu_B(B_{ext}+B_{nuc})/\hbar$ is the electron Larmor frequency. We also use the effective electron g-factor $g$, the Bohr magneton $\mu_B$, and the Planck constant $\hbar$. In the absence of applied current, KR measurements did not show significant spatial diffusion over the same delay range, indicating that drift dominates diffusion in these structures. We extract $v_d$ from these data by taking the slope $\Delta x/\Delta t$ of line cuts along the density plot. As shown in Fig. 2(b), the drift velocity is a linear function of channel current, $i_{channel}$. Separate two-point resistivity measurements indicated no measurable sample heating for the range of currents we used.

On a channel with a MnAs island, we separated the pump and probe beams by 35 μm and focused them on either side of the 30 μm long MnAs island [Fig. 1(c)]. First, we measured KR vs. delay time for various settings of $i_{channel}$ without polarizing the nuclear region as shown in Fig. 3(a). The measured Kerr rotation signal was strongest for delays between 2 to 7 ns after injection, when the maximum amplitude of the spin ensemble emerged from under the MnAs. Next, we polarized the localized nuclear spins with the imprinting beam. To probe the nuclear dynamics, we measured KR at a fixed pump-probe delay indicated by the arrow on Fig. 3(a) for $i_{channel}$ = 0.75 mA as a function of time after blocking the imprinting beam. Figure 3(b) shows how the electron spin projection measured at the probe spot evolved over 20 minutes. We extract the spin rotation under the MnAs by fitting traces of these data at fixed delay to the following expression

$$S_0\left(1 - s_0 \exp\left(\frac{-t}{\tau_N}\right)\right) \cos\left(\Delta\phi_{probe} \exp\left(\frac{-t}{\tau_N}\right)\right), \quad (2)$$

where $\Delta\phi_{probe}$ is the maximum acquired spin rotation, $s_0$ scales the signal due to inhomogeneous nuclear polarization, $t$ is time, and $\tau_N$ is the nuclear spin relaxation time. The evolution of the spin rotation in time is plotted in the inset of Fig. 3(b). These data suggest that $B_{nuc}$ relaxes after active imprinting is turned off, with a timescale on the order of minutes which is consistent with nuclear relaxation times [24]. To confirm that this interaction is due to nuclear spins, we also studied the temperature dependence of this spin rotation at fixed delay, imprinting beam power, and channel current [Fig. 3(c)]. As the temperature increased, we measured less spin rotation over the 20 minute wait time and by 50 K, no spin rotation was measured. Previous studies have observed similar reduction of the FPP effect [19, 20].

Since we place the pump and probe beams near the boundaries of the MnAs island, we can ignore the unpolarized regions and model our experiment as an electron spin ensemble travelling through a uniform polarized nuclear environment. The maximum spin rotation angle at the probe spot is described by

$$\Delta\phi_{probe} = \frac{g\mu_B(B_{ext} + B_{nuc})\Delta t_{nuc}}{h}, \qquad (3)$$

where $\Delta t_{nuc}$ is the traverse time across the region.

This model suggests we can tune the spin rotation by controlling the strength of $B_{nuc}$ at fixed values of $B_{ext}$ and $\Delta t_{nuc}$. In Figure 4(a) we plot the extracted spin rotation vs. the power of the imprinting beam. We observed less rotation at lower power since the reduced presence of injected spin-polarized carriers weakens the FPP effect, resulting in a reduced $B_{nuc}$. The spin rotation saturated near 1 mW, suggesting $B_{nuc}$ was maximized. We calculated this maximum optically-induced value to be 0.4 T using Eq. 3. At even higher powers, sample heating limits the FPP, thus reducing the spin rotation. At optimal power conditions, we measured nearly $5\pi$ radians of nuclear field-induced spin rotation during transport.

The spin rotation also depends on the traverse time of the spins moving through the nuclear region. Using the same method described above, we measured the spin rotation as a function of $\Delta t_{nuc}$ at 8 K and fixed imprinting power. To ensure that a measurable quantity of spins emerged at each delay, we used $i_{channel}$ = 0.75, 1.25, and 1.75 mA for different ranges of $\Delta t_{nuc}$ during the measurement [Fig. 4(c)]. The data show a linear relationship between the spin rotation and traverse time. A fit to this data indicates a slope of $0.86\pi \pm 0.06\pi$ radians/ns.

We repeated our measurements on a second sample [Fig. 1(b)] designed to control the nuclear field strength electrically rather than optically. After processing this sample in an identical manner, an additional metal contact was deposited onto the MnAs to enable electrical biasing of the Schottky junction diode at the MnAs/GaAs interface. Current-voltage characterization of this junction showed expected rectifying behavior as shown by the solid line in Fig. 4(b). For these measurements no imprinting beam was applied; instead, the nuclear spins were polarized by forward-biasing the Schottky junction between the MnAs and GaAs as illustrated in Fig. 1(b). The Schottky current drives electrons from the GaAs to the MnAs inducing spin-dependent reflections at the ferromagnet/semiconductor interface [23]. The resulting spin-polarized electron population at the interface polarizes the nuclear spins through the DNP mechanism [22, 23].

For measurements using electrical polarization, the pump and probe beams were blocked and $i_{channel}$ was set to zero while a voltage was applied to the Schottky junction. After 20 minutes the Schottky voltage was turned off, the pump and probe beams were unblocked, and $i_{channel}$ = 0.75 mA was applied. We observed the onset of spin rotation [open circles in Fig. 4(b)] near the 2 V Schottky turn-on voltage. As the Schottky voltage increased further, the concentration of spin-polarized electrons and resulting nuclear polarization at the MnAs/GaAs interface increased correspondingly. Using Eq. 3, we calculate the maximum electrically-induced value of $B_{nuc}$ to be 0.4 T. Beyond 50 µA, however, we observed a systematic reduction of the measured rotation, suggesting that current-induced heating begins to reduce the strength of the FPP effect.

In summary, we have demonstrated the ability to both electrically and optically manipulate the spin rotation of moving electrons through localized effective magnetic fields in a semiconductor channel. Extensions of these techniques may generate sequenced spin rotation about different axes, potentially also using spatially-localized electric fields, spin-orbit interactions, or pulsed microwaves. Such a device could perform successive operations on moving spins to enable environmental decoupling or quantum metrology of the semiconductor host. Localized control of drifting spins in different materials may also help realize room temperature [26] quantum information processing based on existing transport-based technologies.

This work was supported by ONR MURI under grant N0014-06-1-0428 and by NSF under grants DMR-0801406 and -0801388.

* To whom correspondence should be addressed. Email: awsch@physics.ucsb.edu

**References:**

1. A. Abragam, *Principles of Nuclear Magnetism* (Clarendon Press, Oxford, 1961).

2. J. A. Gupta, R. Knobel, N. Samarth, D. D. Awschalom, *Science* **292**, 2458 (2001).

3. F. H. L. Koppens, C. Buizert, K. J. Tielrooij, I. T. Vink, K. C. Nowack, T. Meunier, L. P. Kouwenhoven, L. M. K. Vandersypen, *Nature* **442**, 766 (2006).

4. F. Jelezko, T. Gaebel, I. Popa, M. Domhan, A. Gruber, J. Wrachtrup, *Phys. Rev. Lett.* **93**, 130501 (2004).

5. J. Berezovsky, M. H. Mikkelsen, N. G. Stoltz, L. A. Coldren, D. D. Awschalom, *Science* **320**, 349 (2008).

6. K. C. Nowack, F. H. L Koppens, Y. V. Nazarov, L. M. K. Vandersypen, *Science* **318**, 1430 (2007).

7. M. A. Nielsen, I. L. Chuang, *Quantum Computation and Quantum Information* (Cambridge Univ. Press, Cambridge, 2002).

8. S. Datta, B. Das, *Appl. Phys. Lett.* **56**, 665 (1990).

9. D. Loss, D. P. DiVincenzo, *Phys. Rev. A.* **57**, 120 (1998).

10. J. M. Kikkawa, D. D. Awschalom, *Phys. Rev. Lett.* **80**, 4313 (1998).

11. J. M. Kikkawa, D. D. Awschalom, *Nature* **397**, 139 (1999).

12. S. A. Crooker, M. Furis, X. Lou, C. Adelmann, D. L. Smith, C. J. Palmstøm, P. A. Crowell, *Science* **309**, 2191 (2005).

13. X. Lou, C. Adelmann, S. A. Crooker, E. S. Garlid, J. Zhang, K. S. M. Reddy, S. D. Flexner, C. J. Palmstrøm, P. A. Crowell, *Nat. Phys.* **3**, 197 (2007).

14. S. A. Crooker, D. L. Smith, *Phys. Rev. Lett.* **94**, 236601 (2005).

15. T. Koga, Y. Sekine, J. Nitta, *Phys. Rev. B* **74**, 041302(R) (2006).

16. I. Appelbaum, B. Huang, D. J. Monsma, *Nature* **447**, 295 (2007).

17. N. Tombros, C. Jozsa, M. Popinciuc, H. T. Jonkman, B. J. van Wees, *Nature* **448**, 571 (2007).

18. Y. K. Kato, R. C. Myers, A. C. Gossard, D. D. Awschalom, *Appl. Phys. Lett.* **86**, 162107 (2005).

19. J. Stephens, R. K. Kawakami, J. Berezovsky, M. Hanson, D. P. Shepherd, A. C. Gossard, D. D. Awschalom, *Phys. Rev. B* **68**, 41307(R) (2003).

20. R.K. Kawakami, Y. Kato, M. Hanson, I. Malajovich, J. M. Stephens, E. Johnston-Halperin, G. Salis, A. C. Gossard, D. D. Awschalom, *Science* **294**, 131 (2001).

21. G. E. W. Bauer, A. Brataas, Y. Tserkovnyak, B. I. Halperin, M. Zwierzycki, P. J. Kelly, *Phys. Rev. Lett.* **92**, 126601 (2004).

22. J. Stephens, J. Berezovsky, J. P. McGuire, L. J. Sham, A. C. Gossard, D. D. Awschalom, *Phys. Rev. Lett.* **93**, 097602 (2004).

23. C. Ciuti, J. P. McGuire, and L. J. Sham, *Phys. Rev. Lett.* **89**, 156601 (2002).

24. F. Meier, B. P. Zakharchenya, Eds., *Optical Orientation: Modern Problems in Condensed Matter* (North Holland, Amsterdam, 1984).

25. R. J. Epstein, I. Malajovich, R. K. Kawakami, Y. Chye, M. Hanson, P. M. Petroff, A. C. Gossard, D. D. Awschalom, *Phys. Rev. B* **65**, 121202 (2002).

26. N. P. Stern, S. Ghosh, G. Xiang, M. Zhu, N. Samarth, D. D. Awschalom, *Phys. Rev. Lett.* **97**, 126603 (2006).

**Figures**

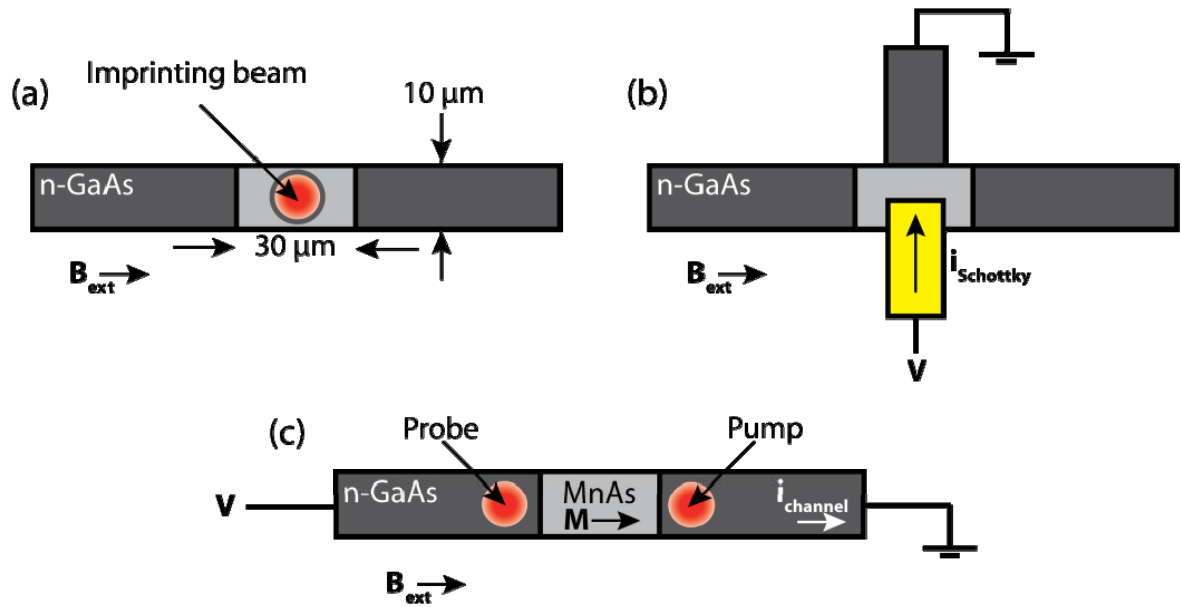

Figure 1. Nuclear spins are polarized at the MnAs/GaAs interface using (a) an optical imprinting beam, or (b) a forward-biased Schottky junction. In both cases no current flows through the channel during the imprinting process. (c) Non-local TRKR measurement geometry for both experiments.

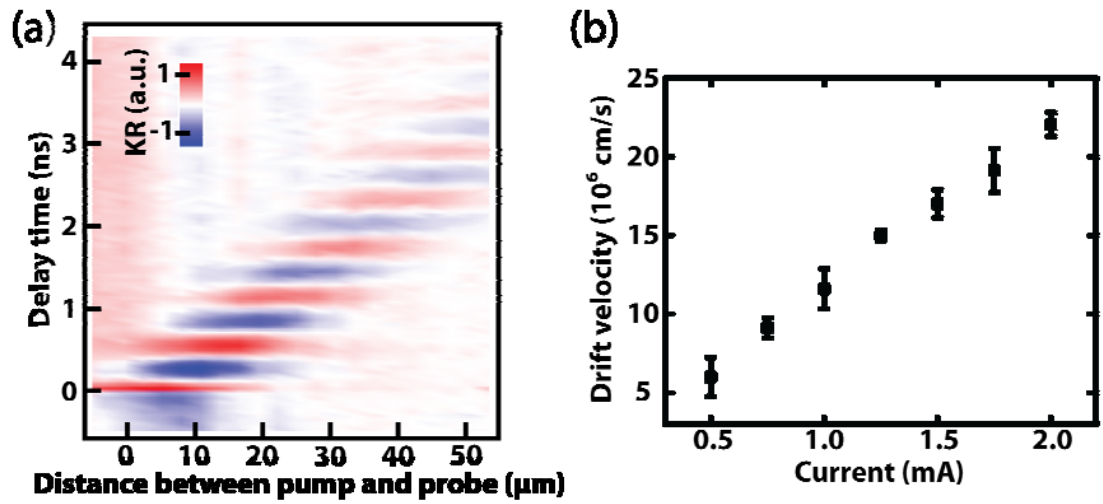

Figure 2. (a) Time-resolved KR vs delay time in a control GaAs channel (with no MnAs) as the pump beam is rastered away from the probe beam along the channel with $B_{ext}$ = 0.2850 T, $T$ = 8 K, and $i_{channel}$ = 1 mA. (b) Drift velocity extracted from rastered TRKR scans as a function of channel current.

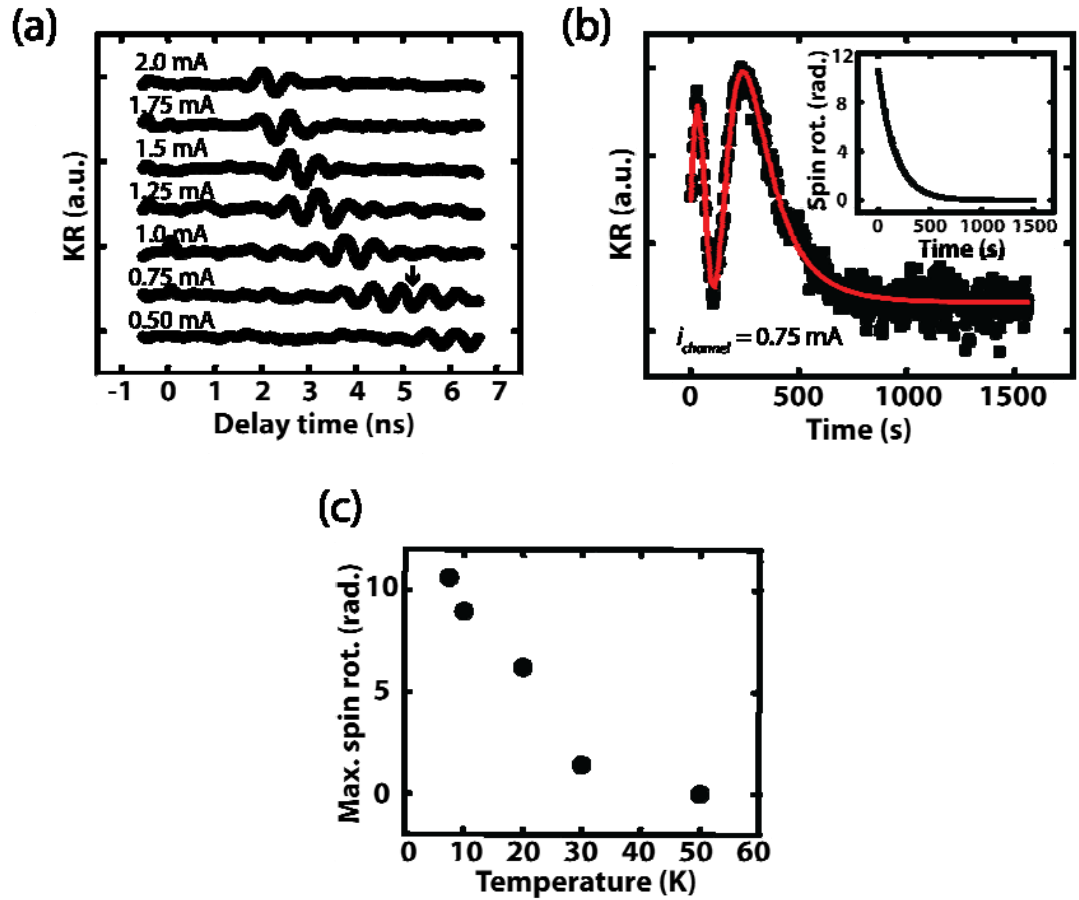

Figure 3. (a) Non-local TRKR delay scans at fixed pump/probe separation of 35 μm for specified channel currents between 0.5 to 2.0 mA at $T$ = 8 K and $B_{ext}$ = 0.2850 T. These data are taken in a patterned MnAs/GaAs device using the scheme shown in Fig. 1(c). (b) Non-local TRKR vs. time data at fixed delay indicated by arrow in (a) for $i_{channel}$ = 0.75 mA. Data is taken after optically polarizing the nuclei [Fig. 1(a)]. Solid line shows a fit to Eq. 2. Inset: Time dependence of the nuclear-induced spin rotation measured in (b), extracted from Eq. 2. (c) Maximum rotation of the electron spin projection as a function of temperature. Error bars are comparable to the size of the data points.

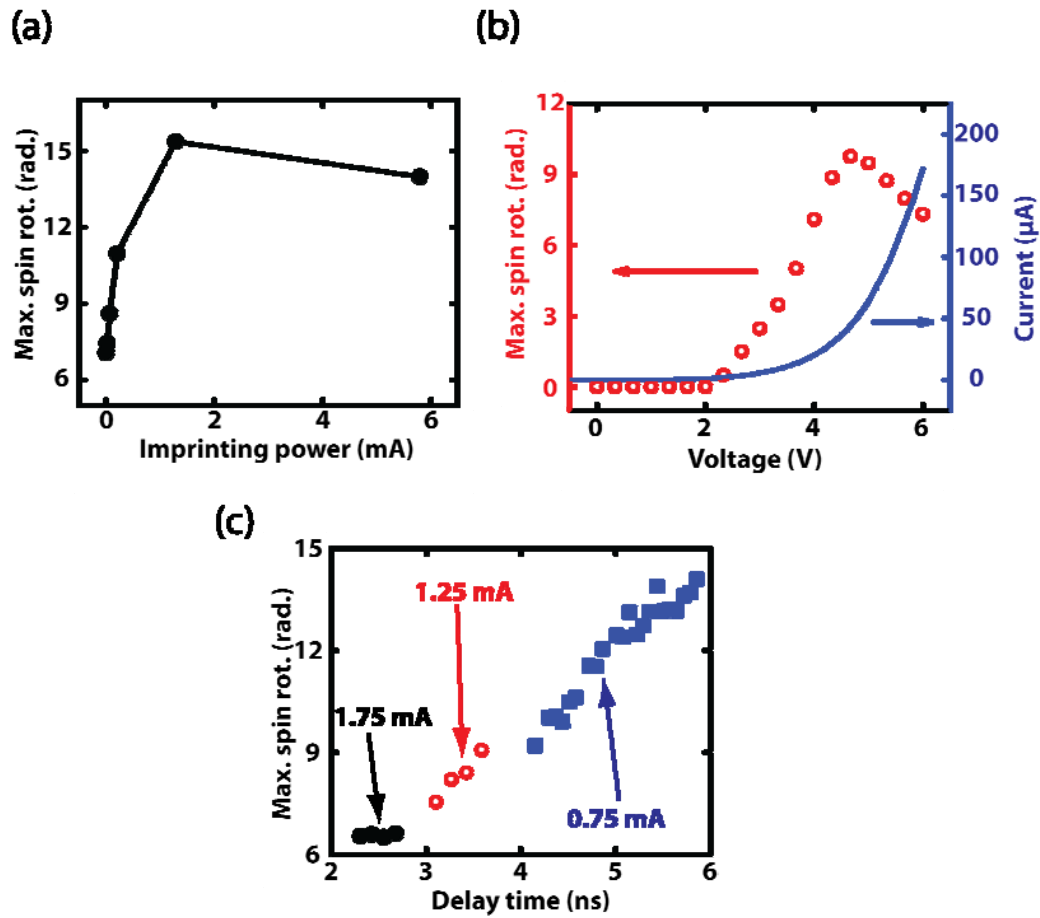

Figure 4. (a) Maximum spin rotation measured as a function of the imprinting beam power at $T$ = 8 K, $\Delta t$ = 6 ns, and $i_{channel}$ = 0.75 mA. (b) Schottky current and maximum spin rotation measured vs. forward-bias Schottky voltage at $T$ = 8 K, $\Delta t$ = 4.5 ns, and $i_{channel}$ = 0.75 mA. (c) Maximum spin rotation measured as a function of delay time for $i_{channel}$ = 0.75, 1.25, and 1.75 mA. Error bars are comparable to the size of the data points.